\documentstyle[editedbook,epsfig,psfig,epsf]{mq}

\def\etal{{\em et al.} }

\def\cm2{cm$^2$ }
\def\se1{s$^{-1}$ }

\begin{opening}
\title{3-D GRMHD Simulations of Generating Jets}
\author{K.-I. Nishikawa$^1$, S. Koide$^2$, K. Shibata$^3$, T. Kudoh$^{4,5}$,
H. Sol$^6$} 
\institute{$^1$ National Space Science and Technology Center,
   320 Sparkman Drive, SD 50,  Huntsville, AL 35805, USA.\\
$^2$ Toyama University, Faculty of Engineering, 3190 Gofuku, Toyama 930-8555,
   Japan.\\
$^3$ Kyoto University, Kwasan Observatory, Yamashina, Kyoto 607-8471, Japan.\\
$^4$ National Astronomical Observatory, Mitaka, Tokyo 181-8588, Japan.\\
$^5$  Department of Physics and Astronomy, University of Western Ontario
   London, Ontario N6A 3K7, Canada.\\
$^6$ LUTH, Observatoire de Paris-Meudon, 92195 Meudon Cedox, France.}
\author{J. P. Hughes$^7$, G. Richardson$^1$, R. Preece$^1$ 
P. Hardee$^8$}
\institute{$^7$ Department of Physics and Astronomy, Rutgers, The State 
   University of New jersey, Piscataway, NJ 08854-8019, USA.\\
$^8$ Department of Physics and Astronomy, The University of Alabama,
  Gallalee Hall, Tuscaloosa, AL 35487, USA.}
\end{opening}

\runningtitle{Workshop Proceedings}
\runningauthor{Nishikawa et al.}

\begin{document}
\vspace{-0.5cm}
\begin{abstract}
{\small We have performed a first fully 3-D GRMHD simulation with 
Schwarzschild black hole with a free falling corona. The initial simulation 
results show that a jet is created as in the previous simulations using the
axisymmetric geometry with the mirror symmetry at the equator. However, the 
time to generate the jet is longer than in the 2-D axisymmetric simulations. 
We expect that due to
the additional freedom in the azimuthal dimension without axisymmetry
with respect to the $z$ axis and reflection symmetry with respect to the
equatorial plane, the dynamics of jet formation can be modified. Further
simulations are required for the study of instabilities along the azimuthal
direction such as accretion-eject instability}
\end{abstract}

\section{Introduction}

   Relativistic jets have been observed in active galactic nuclei (AGNs)
and microquasars in our Galaxy, and it is believed that they originate from 
the regions near accreting black holes or neutron stars \cite{mku01}. 
To investigate the dynamics of accretion disks and the associated jet 
formation, we use a full 3-D GRMHD code.  One of the most promising models 
for jet formation is the magnetic-acceleration model \cite{bp82}. 
The magnetic-acceleration mechanism has
been proposed not only for AGN jets, but also for protostellar jets (see
\cite{mku01}). 

   Recently, Koide, Shibata, \& Kudoh (1999) \cite{ksk99} have investigated 
the dynamics
of an accretion disk initially threaded by a uniform poloidal magnetic field
in a non-rotating corona (either in a state of steady fall or in hydrostatic
equilibrium) around a non-rotating black hole. The numerical results show
that the disk loses angular momentum by magnetic braking, then falls into the
black hole.  The infalling motion of the disk, which is faster than in
the non-relativistic case because of the general-relativistic
effect below $3 r_{\rm S}$ ($r_{\rm S}$ is the Schwarzschild radius),
is strongly decelerated at the shock formed by the centrifugal force
around $r = 2 r_{\rm S}$ by the rotation
of the disk. Plasmas near the shock are accelerated by the ${\bf J}
\times {\bf B}$ force, which forms bipolar relativistic jets. Inside this
{\it magnetically driven jet}, the gradient of gas pressure
also generates a jet
over the shock region ({\it gas-pressure-driven jet}).
This {\it two-layered jet structure} is formed both in a hydrostatic corona
and in a steady-state falling corona. Koide et al.~(2000) \cite{kmsk00} have 
also developed a new GRMHD code in Kerr geometry and have found that,
with a rapidly rotating ($a = 0.95$) black-hole magnetosphere,
the maximum velocity of the jet is 0.9 c and its terminal velocity 0.85 c.
All of the previous 2-D GRMHD simulations described here were made
assuming axisymmetry with respect
to the $z$-axis and mirror symmetry with respect to the plane $z = 0$;
the axisymmetric assumption suppressed the azimuthal instabilities.

\section{3-D GRMHD Simulations: Equations and Numerical Techniques}

Our basic equations are those of Maxwell for the fields
and a set of general-relativistic equations representing the plasma, namely
the equations of conservation of mass, momentum, and energy for a
single-component conducting fluid \cite{wein72,tpm86,ksk99}.

   Our simulations employ the equations of general-relativistic
MHD together with the Simplified Total-Variation-Diminishing (STVD) method.
The STVD method was developed by Davis (1984) \cite{davis84} for studying
violent phenomena such as shocks \cite{nishikawa97,ksk99}.
This method is as Lax-Wendoroff's method with additional diffusion term.
We have checked that this method respects
the energy-conservation law and its propagation properties \cite{ksk99}.

\section{Initial 3-D GRMHD simulations with a Schwarzschild
    black hole}

In order to investigate how accretion disks near
black holes evolve under the influence of accretion instabilities
such as the magnetorotational instability \cite{bh98} 
and accretion-ejection instability (AEI) \cite{tagger99}, 
the use of a fully 3-D GRMHD is essential.

\subsection{Initial and boundary conditions}
In the assumed initial state, the simulation region is divided into two parts:
a background corona around a black hole, and an accretion disk (Fig. 1a). 
The coronal plasma is set in a state of transonic free-fall flow,
as in the case of the transonic flows with $\Gamma = 5/3$ and $H=1.3$; here
the sonic point is located at $r=1.6 r_{\rm S}$.
The Keplerian disk in the corona is set in the following way.
The disk region is located at $ r > r_{\rm D} \equiv 3r_{\rm S},
| {\rm cos} \theta | < \delta =1/8$. Here the density is 100 times that of the
background corona (Fig. 1a), while the orbital velocity is relativistic
and purely azimuthal: $v_\phi = v_{\rm K} \equiv c/[2(r/r_{\rm S} -1)]^{1/2}$.
(Note that this equation reduces to the Newtonian
Keplerian velocity $v_\phi =\root \of {GM/r}$ in
the non-relativistic limit $r_{\rm S}/r \ll 1$).
The pressure of both the corona and the disk are assumed equal
to that of the transonic solution.  The initial conditions for the entire
plasma around the black hole are:
$\rho = \rho _{\rm ffc} +\rho _{\rm dis}$

\begin{equation}
\rho _{\rm dis} = \left \{  \begin{array}{cc}
100 \rho _{\rm ffc} &
\verb!   ! ( r > r_{\rm D} \verb!  ! {\rm and} \verb!  !
|{\rm cot} \theta| < \delta )\\ 0                  &
\verb!   ! ( r \leq r_{\rm D} \verb!  ! {\rm or} \verb!  !
|{\rm cot} \theta| \geq \delta )
\end{array} \right .
\end{equation}

\begin{equation}
(v_r, v_\theta , v_\phi) = \left \{  \begin{array}{cc}
(0, 0, v_{\rm K}) &
\verb!   ! ( r > r_{\rm D} \verb!  ! {\rm and} \verb!  !
|{\rm cot} \theta| < \delta )\\ (-v_{\rm ffc}, 0, 0)        &
\verb!   ! ( r \leq r_{\rm D} \verb!  ! {\rm or} \verb!  !
|{\rm cot} \theta| \geq \delta )
\end{array} \right .
\end{equation}
where we set $\delta = 0.125$; the
smoothing length is $0.3 r_{\rm S}$.

In addition, there is a magnetic field crossing
the accretion disk perpendicularly.
We set it to the Wald solution \cite{wald74}, 
which represents the uniform magnetic field
around a Kerr black hole:
$B_r = B_0 {\rm cos} \theta$,
$B_{\theta } = - \alpha B_0 {\rm sin} \theta $ (where
$\alpha$ is the lapse function, $\alpha = (1 - r_{\rm S}/r)^{1/2}$).
At the inner edge of the accretion disk,
the proper Alfv\'{e}n velocity is $v_{\rm A} = 0.015c$
in a typical case with $B_0 =0.3 \, \root \of {\rho _0c^2}$,
where the
Alfv\'{e}n velocity in the fiducial observer,
$v_{\rm A} \equiv B/{\root \of {\rho + [\Gamma p/(\Gamma -1)+B^2]/c^2}}$.
The plasma beta of the corona at $r=3r_{\rm S}$
is $\beta \equiv p/B^2=1.40$.
The simulation is performed in the region
$1.1 r_{\rm S} \leq r \leq 20 r_{\rm S}$, $0 \leq \theta \leq
\pi$, $0 \leq \phi \leq 2\pi$ with $100 \times 120 \times 60$ meshes.
The effective linear
mesh widths at $r=1.1r_{\rm S}$ and at
$r=20r_{\rm S}$
are $5.38 \times 10^{-3} r_{\rm S}$ and $0.97 r_{\rm S}$,
respectively, while the angular spacings along the polar and azimuthal
directions are $5.2 \times 10^{-2}$ rad.
A radiative boundary condition is imposed at $r=1.1 r_{\rm S}$ and at
$r=20 r_{\rm S}$.
The computations were made on an ORIGIN 2000 computer
with 0.898 GB internal memory, and they used
about 47 hours of CPU time for 10000 time steps
with $100 \times 120 \times 60$ meshes.

\begin{figure}[htb]
\epsfig{file=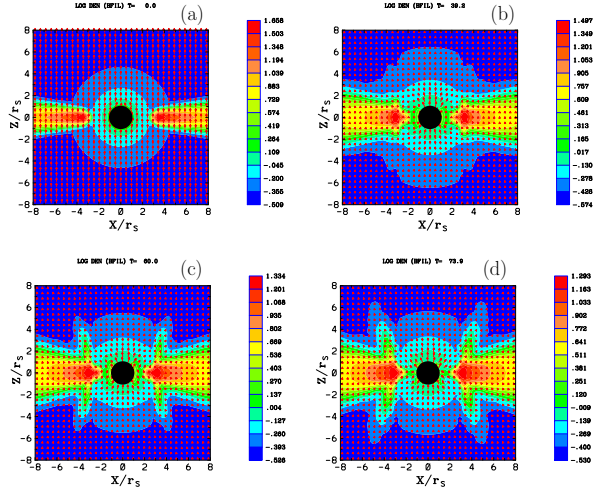,width=7.8cm}
\centering
\caption{For the fully 3-D simulation,these panels present the time
evolution of the proper mass density with the magnetic field ($B_{x}, B_{z}$)
in a transonic free-fall (steady-state falling) corona with an initially
uniform magnetic field, at $t=0.0\tau _{\rm S}$ (a),  $t=39.3\tau _{\rm S}$ (b),$t=60.0\tau _{\rm S}$ (c), and $t=73.9 \tau_{\rm S}$ (d). The jet is formed
around $r = 4.5 r_{\rm S}$ as in the 2-D simulation.}
\end{figure}

\subsection{Simulation results}

Figure 1 shows the evolution of 3-D simulation
performed in the region
$1.1 r_{\rm S} \le r \le 20 r_{\rm S}$,  $0 \le \theta \le \pi$,
and $0 \le \phi \le 2\pi$ with $100 \times 120 \times 60$ meshes.
The parameters used in this simulation are the same as those of the
axisymmetric simulations shown in Fig. 6 of \cite{ksk99}. 
In this figure, the colored shading shows the proper mass density on
logarithmic scales; the vector plots show the magnetic field.

The black circle represents the black hole. Figures 1a presents
the initial conditions, which are the same as in the 2-D simulations 
\cite{ksk99}. 
At  $t= 39.2 \tau_{\rm S}$ (Fig. 1b) comparing with
Fig. 6c ($t= 40.0 \tau_{\rm S}$) in \cite{ksk99} 
the jet is less
generated. At $t= 60.0 \tau_{\rm S}$ as shown in Fig. 1c, the jet is generated
as 2-D simulation
at the earlier time ($t= 40.0 \tau_{\rm S}$). At $t= 73.9 \tau _{\rm S}$,
the jet is clearly created around $r = 4.5 r_{\rm S}$, which is shown by the
enhanced density (Fig. 1d). As in the 2-D simulations
the jet is generated in a hollowed cylindrical form.

The delay of jet formation seems to be due to reduction of shock formation at
$r = 2 r_{\rm S}$ caused by the additional freedom in the azimuthal
direction. Further investigation will be reported elsewhere.

\section{Discussion}

   Recently, review articles on magnetohydrodynamic production of
relativistic jets have reported unsolved questions related to jet
formation and its mechanisms which determine the velocity of jets and
time variations of jet flux \cite{mku01,bld01,meier02}.

  This simulation result is initial and we will perform more simulations
and investigate effects of the third dimension. 
This simulation study will 
be extended to understand the different (high/soft and low/hard) states
\cite{fender01} 
and different combination of accreting rate $\dot{m} \equiv
\dot{M}/\dot{M}_{Edd}$ and the angular momentum $j \equiv J/J_{\rm max}$
\cite{meier02}. 
Further results will be reported elsewhere.

\section*{Acknowledgments}
K.N. is partially supported by NSF ATM 9730230,
ATM-9870072, ATM-0100997, and INT-9981508. The simulations have been performed
on ORIGIN 2000 at NCSA which is supported by NSF.

\end{document}